\documentclass[aps,prl,twocolumn,superscriptaddress]{revtex4-2}

\usepackage{graphicx}
\usepackage{amssymb}
\usepackage{braket}
\usepackage{comment}
\usepackage[colorinlistoftodos]{todonotes}

\newcommand{\tripletSone}{{^3\hspace{-.03in}S_1}}
\newcommand{\tripletPzero}{{^3\hspace{-.03in}P_0}}

\newcommand{\tripletPone}{{^3\hspace{-.03in}P_1}}

\begin{document}

\title{Realizing Su-Schrieffer-Heeger topological edge states in Rydberg-atom synthetic dimensions}


\author{S. K. Kanungo}
\affiliation{Department of Physics and Astronomy, Rice University, Houston, TX  77005-1892, USA}
\affiliation{Rice Center for Quantum Materials, Rice University, Houston, TX  77005-1892, USA}
\author{J. D. Whalen}
\affiliation{Department of Physics and Astronomy, Rice University, Houston, TX  77005-1892, USA}
\affiliation{Rice Center for Quantum Materials, Rice University, Houston, TX  77005-1892, USA}

\author{Y. Lu}
\affiliation{Department of Physics and Astronomy, Rice University, Houston, TX  77005-1892, USA}
\affiliation{Rice Center for Quantum Materials, Rice University, Houston, TX  77005-1892, USA}

\author{M. Yuan}
\thanks{Current address: Pritzker School of Molecular Engineering, University of Chicago, Chicago, Illinois 60637, USA}
\affiliation{Department of Physics and Astronomy, Rice University, Houston, TX  77005-1892, USA}
\affiliation{Rice Center for Quantum Materials, Rice University, Houston, TX  77005-1892, USA}
\affiliation{School of the Gifted Young, University of Science and Technology of China, Hefei 230026, China}

\author{S. Dasgupta}
\affiliation{Department of Physics and Astronomy, Rice University, Houston, TX  77005-1892, USA}
\affiliation{Rice Center for Quantum Materials, Rice University, Houston, TX  77005-1892, USA}

\author{F. B. Dunning}
\affiliation{Department of Physics and Astronomy, Rice University, Houston, TX  77005-1892, USA}

\author{K. R. A. Hazzard}
\author{T. C. Killian}

\affiliation{Department of Physics and Astronomy, Rice University, Houston, TX  77005-1892, USA}
\affiliation{Rice Center for Quantum Materials, Rice University, Houston, TX  77005-1892, USA}

\begin{abstract}
We demonstrate a platform for synthetic dimensions based on coupled Rydberg levels in ultracold atoms, and we implement the single-particle  Su-Schrieffer-Heeger (SSH) Hamiltonian. Rydberg levels  are interpreted as synthetic lattice sites, with tunneling  introduced through resonant millimeter-wave couplings. Tunneling amplitudes are controlled through the millimeter-wave amplitudes, and on-site potentials are controlled through detunings of the millimeter waves from resonance. Using alternating weak and strong tunneling with weak tunneling to edge lattice sites, we attain a configuration with symmetry-protected topological edge states. The band structure is probed through optical excitation to the  Rydberg levels from the ground state, which reveals topological edge states at zero energy. We verify that edge-state energies are robust to perturbation of tunneling-rates, which preserves chiral symmetry, but can be shifted by the introduction of on-site potentials.

\end{abstract}

\maketitle

A synthetic dimension  \cite{boada2012,opr19} is a degree of freedom encoded into a set of internal or external states  that can mimic  a real-space lattice potential.
Synthetic dimensions  are powerful tools for quantum simulation, opening exciting possibilities such as the realization of higher dimensional systems~\cite{tsomokos2010, boada2012, jukic2013}, non-trivial real space \cite{Boada_2015} and band structure \cite{bansil2016,qi2011} topologies, and artificial gauge fields \cite{celi2014,ang18}.
Experiments have utilized various degrees of freedom \cite{opr19} to create synthetic dimensions, such as motional~\cite{An2017,anPRL17}, spin~\cite{mancini1510,celi2014,anisimovas16,wall16}, and rotational~\cite{Flo15} levels of atoms and molecules, and frequency modes, spatial modes, and arrival times in photonic systems~\cite{opr19}. 

Prominent demonstrations of atomic synthetic dimensions include observation of artificial gauge fields, spin-orbit coupling, and chiral edge states in  Raman-coupled ground magnetic sublevels~\cite{celi2014,mpc15,sla15} or single-photon-coupled electronic orbitals~\cite{lcd16,kbb17} grafted onto motion in a real 1D optical lattice. A synthetic dimension can also be formed by discrete motional states~\cite{pog17}, such as
free-particle momentum states coupled with momentum-changing two-photon Bragg transitions~\cite{gad15,mag16PRA}. The latter has been used to observe Anderson localization~\cite{mad18}, artificial gauge fields~\cite{amg18}, and topological states~\cite{mag16NatComm,xgx19}. 

\begin{figure}
  \includegraphics[scale = 0.5]{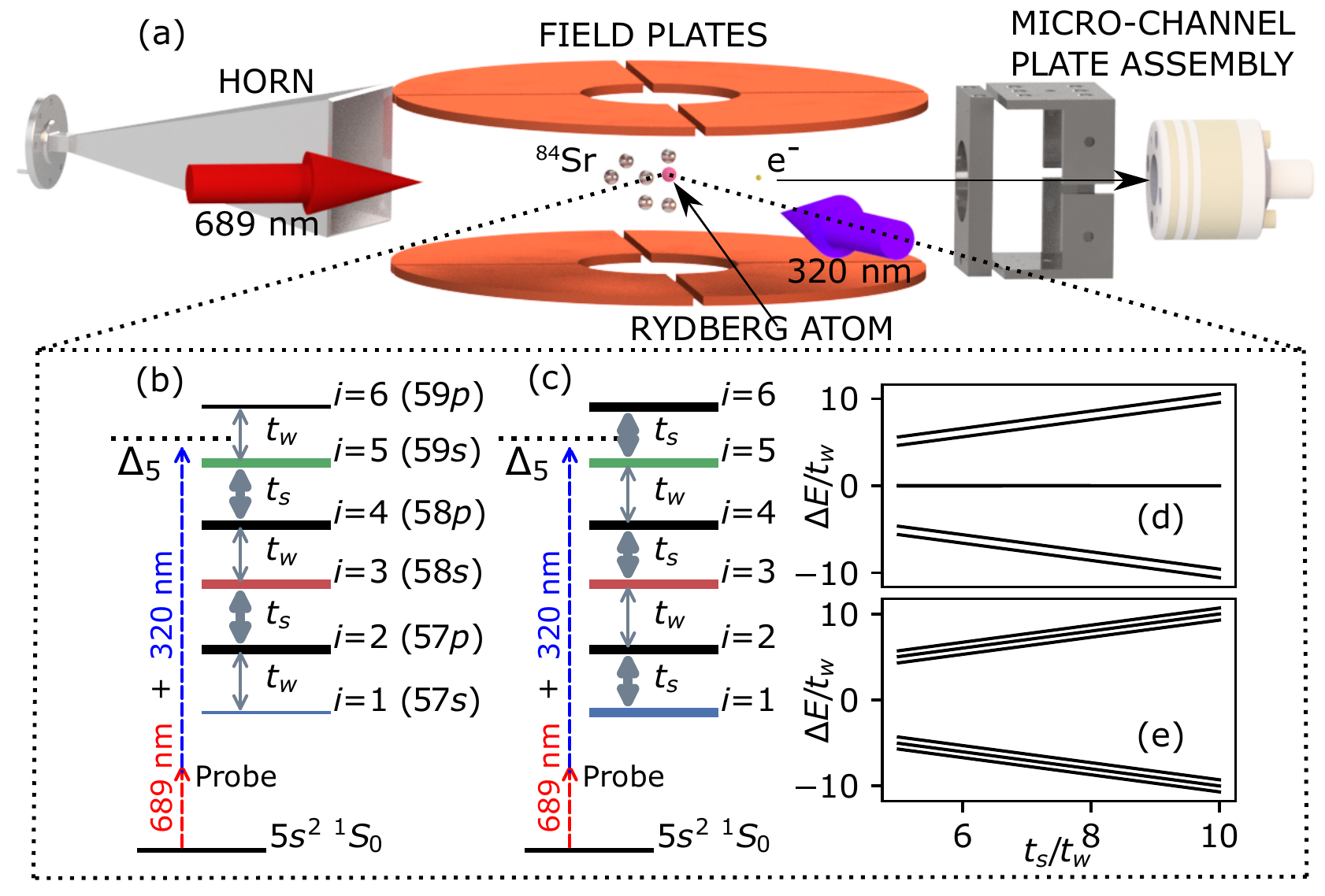}
 \caption{\label{fig:Level_diagram} (a) Experimental schematic. (b) Configurations with  and (c) without TPS 
 of the SSH model using six Rydberg levels of $^{84}$Sr. Double-headed grey arrows denote near-resonant millimeter-wave couplings, which induce tunneling between sites of the synthetic lattice, and thicker lines correspond to faster tunneling. Dashed lines show  two-photon excitation to a Rydberg level of interest. (d,e) show band structure for (b) and (c) respectively vs. the ratio of tunneling amplitudes, \ $t_s/t_w$. The site-numbering convention is given in (b), with odd numbers corresponding to $ns$ states.}
\end{figure}

Here we harness Rydberg levels of $^{84}$Sr to realize a synthetic lattice for studying quantum matter. Millimeter-wave couplings between Rydberg states introduce  tunneling between synthetic lattice sites. This scheme was suggested in \cite{opr19}, and it is similar to a proposal for synthetic dimensions based on molecular rotational levels~\cite{sgh18,stw19,bgb20}. It allows for control of the connectivity, tunneling rates, and on-site potentials, and creation of a broad range of synthetic dimensional systems. 

To demonstrate this technique, we realize the Su–Schrieffer–Heeger (SSH) model \cite{SSH1979}  with six  lattice sites  formed with  three $5sns$\,$\tripletSone(m=1)$ ($\equiv ns$, sites $i=1,3,5$, with $57s$ mapped to $i=1$) and three $5snp$\,$\tripletPzero$ ($\equiv np$, $i=2,4,6$) levels [Fig.\ \ref{fig:Level_diagram}(b)], and study its topologically protected edge states (TPS) and their robustness to disorder.
The SSH model describes a linear conjugated polymer, such as polyacetylene, with alternating weak and strong tunneling. The configuration with weak tunneling to edge sites possesses doubly degenerate TPS with energy centered in the gap between bulk states. TPS energies are robust against perturbations respecting the chiral symmetry of the tunneling pattern~\cite{cts16,cds19}, as observed  in many systems~\cite{atala2013, mag16NatComm, stjean2017, sylvain2019}. 

The Hamiltonian  realized is
\begin{equation}
\label{eq:SSH}
\hat{\text{H}}_{\text{lattice}} = \sum_{i=1}^{5}(-ht_{i,i+1}\ket{i}\!\bra{i+1} + \text{h.c.}) +\sum_{i=1}^{6}h\delta_i\ket{i}\!\bra{i},
\end{equation}
where $t_{i,i+1}$ are the tunneling amplitudes and $\delta_i$ are on-site potentials set respectively by amplitudes and detunings of the millimeter-wave couplings, and $h$ is Planck's constant. To obtain Eq.~(\ref{eq:SSH}), we have neglected counter-rotating terms in the millimeter-wave couplings and  transformed into a rotating frame. The kets $\ket{i}$ correspond to the unperturbed Rydberg levels of $^{84}$Sr, up to a time-dependent phase arising from the transformation.  $\delta_i=0$ yields the SSH model, and the  configuration with TPS 
has $t_{i,i+1}$ = $t_{w}$ ($t_{s}$) for $i =1, 3, 5 (2, 4)$ and $t_{w} < t_s$. For the configuration without TPS, 
the weak and strong couplings are exchanged.

The essential elements of the present apparatus are shown in Fig. \ref{fig:Level_diagram}(a). We trap  $10^5$ $^{84}$Sr atoms in an optical dipole trap at a peak density of about $\sim 10^{11}\,\text{cm}^{-3}$ and a temperature of $T= 2\, \mu$K. The laser cooling and trapping of $^{84}$Sr has been described in detail elsewhere \cite{ssk14,mmy09}. 

Before the first excitation cycle, millimeter waves are switched on to  provide  $ns{(m=1)}-np$ and $np-ns{(m=1)}$ coupling for three different $n$'s as shown in Figure \ref{fig:Level_diagram}(b). Millimeter-wave frequencies are generated by combining outputs of five RF synthesizers ($<6$\,GHz) and mixing the result with a 16 GHz local oscillator. A K-band horn antenna rejects the lower sidebands and directs the power in the upper sidebands to the atoms. The coupling strengths can be varied by varying the low-frequency-synthesizer output powers. Here, all data is taken with $t_w=100$\,kHz. 
Each coupling is calibrated using the Autler-Townes splitting \cite{ato55} in a two-level configuration, which is equal to the coupling Rabi frequency  $\Omega=2t$. A 4 Gauss magnetic field splits the $ns$ magnetic sublevels by 11\,MHz, which is large compared to tunneling rates. AC stark shifts are experimentally determined, and the $\delta_i$ in Eq.~\ref{eq:SSH} are relative to the Stark-shifted Rydberg levels.

To populate and probe the synthetic space, the $^{84}$Sr ground state is coupled to the  Rydberg levels via two-photon excitation with an intermediate detuning of +80 MHz from the 5s5p $\tripletPone$ level \cite{dad15,csw18,wkd19}, applied in a $5\,\mu$s pulse.  The laser polarizations selects excitation to  $ns(m=1)$ levels. 
 Immediately after excitation, Rydberg populations are detected using selective field ionization (SFI) \cite{gal94}, for which purpose an electric field  of the form $E(t) = E_p(1-e^{-t/\tau})$  is applied, with $E_p=49$ V/cm and $\tau= 6.5\,\mu$s. An  atom in level $n\ell$ ionizes at a field  given by $\sim 1/[16({n-\alpha_{\ell}})^4$], where $\alpha_{0}=3.371$ and $\alpha_{1}=2.887$ \cite{vjp12} are the quantum defects of the $ns$ and $np$ states respectively. Liberated electrons are detected by a micro-channel plate, and the Rydberg level, or occupied synthetic-lattice site, can be determined from the arrival time of the electron. With the current experimental resolution, arrival times for states $np$ and $(n+1)s$ are unresolved. Approximately $10^4$ excitation cycles are performed per sample at a 4 kHz repetition rate, and the two-photon drive is weak enough that either zero or one atom is excited to the Rydberg manifold each cycle.


To probe the lattice band structure, the two-photon excitation laser is tuned, with detuning $\Delta_{ i_{\text{pr}}}$, near the energy of one of the unperturbed Rydberg levels ($\ket{ i_{\text{pr}}}$). Neglecting far off-resonant terms, the Hamiltonian for the entire system can be written as:
\begin{equation}
\hat{\text{H}} = \frac{h\Omega_{ i_{\text{pr}}}}{2}\ket{g}\!\bra{ i_{\text{pr}}}e^{i2\pi\Delta_{ i_{\text{pr}}}t} + \text{h.c.} + \hat{\text{H}}_{\text{lattice}},
\label{eq:Hamiltonian}
\end{equation}
where $\Omega_{ i_{\text{pr}}}$ denotes the effective two-photon Rabi frequency, which vanishes for even $ i_{\text{pr}}$ ($np$ levels),  and $\ket{g}$ is the ground state vector  in  the frame rotating at the frequency difference of the $\ket{i_{\text{pr}}}$ and $\ket{g}$ levels. The Rydberg excitation rate before convolving with instrumental linewidth is
\begin{equation}
\label{eq:FGR}
\Gamma(\Delta_{ i_{\text{pr}}}) =  \pi^2{\Omega_{ i_{\text{pr}}}^2}\sum_{\beta} |\braket{\beta| i_{\text{pr}}}|^2\delta(\Delta_{ i_{\text{pr}}}-\epsilon_{\beta}/h),
\end{equation}
where $\ket{\beta}$ and and  $\epsilon_{\beta}$ are the eigenstates and eigenenergies of $\hat{\text{H}}_{\text{lattice}}$. The collection of spectra, with each spectrum arising from coupling $\ket{g}$ to a different $\ket{ i_{\text{pr}}}$,  complement each other to provide a  characterization of the band structure and decomposition of the eigenstates because the spectral contribution from each eigenstate is proportional to its overlap with the unperturbed Rydberg level corresponding to the lattice site $ i_{\text{pr}}$. 
\begin{figure}[ht]
  \includegraphics[scale = 0.75]{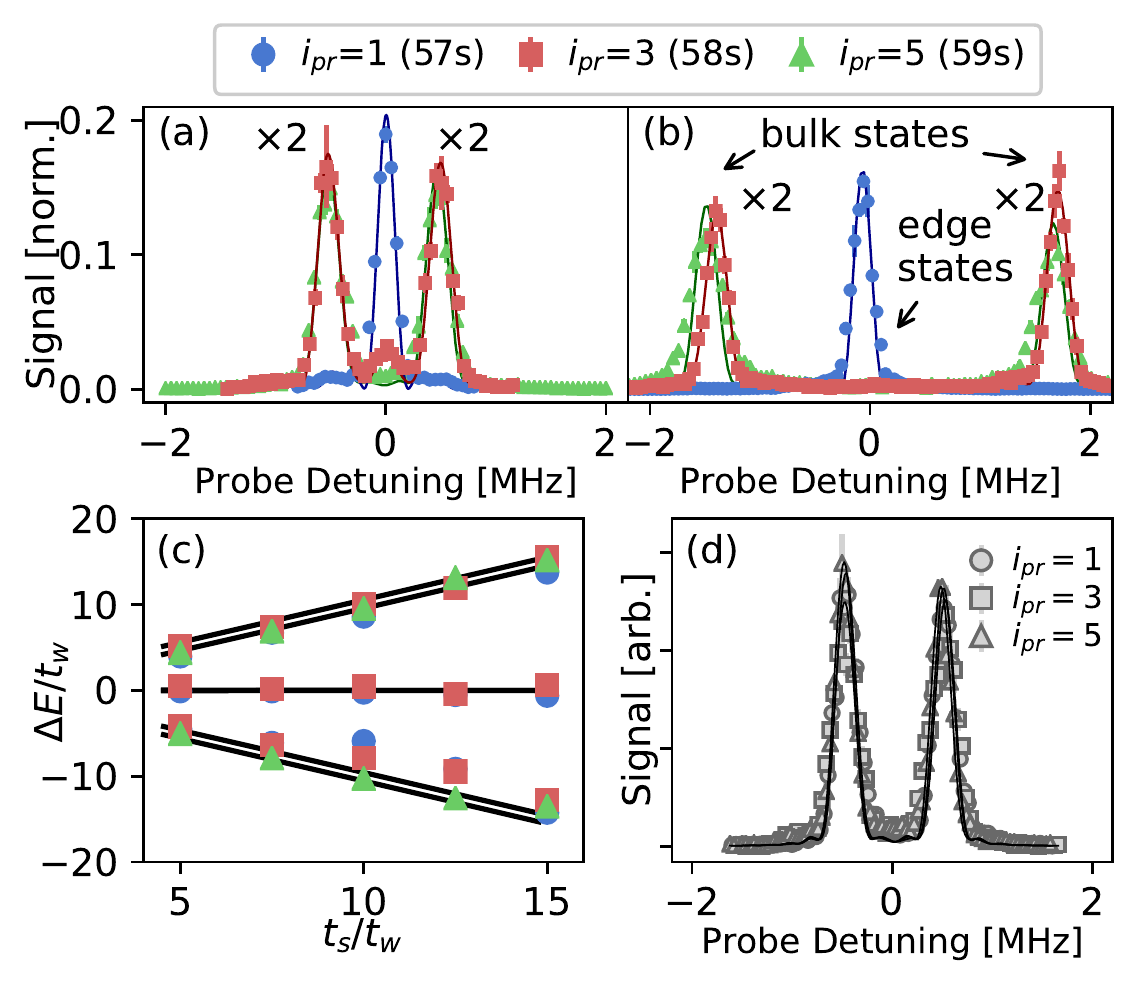}
 \caption{\label{fig:Gallery} (a,b) Rydberg excitation spectra when coupling to $ i_{\text{pr}}=1$($57s$), $ i_{\text{pr}}=3$(58s), and $ i_{\text{pr}}=5$(59s) for $t_s/t_w$=5 (a) and 15 (b) in the  configuration with TPS. Probe detuning ($\Delta_{ i_{\text{pr}}}$) is from the undressed Rydberg level. $ i_{\text{pr}}=3,5$ spectra have been multiplied by a factor of 2 for clarity.  (c) Peak positions in spectra such as in (a,b), giving the bulk and edge state energies versus $t_s/t_w$. (d) Spectra  at $t_s/t_w=5$ for the configuration without TPS.}
\end{figure}

 Figure \ref{fig:Gallery}(a,b) shows spectra for the configuration with TPS 
 and  $\delta_i=0$ as a function of probe-laser detuning near each of the unperturbed Rydberg $ns$ levels (odd $ i_{\text{pr}}$).  Each spectrum is normalized by the total signal for its $i_{\text{pr}}$, and $t_s$ is varied from 0.5\,MHz to 1.5\,MHz. Principal features of the SSH model are readily discerned. Edge states appear at detuning $\Delta_{ i_{\text{pr}}}=0$ in a gap of width $\sim 2t_s$ between bulk states. 
The edge-state signal  is large for probe detuning near the $57s$ level ($ i_{\text{pr}}=1$), small for the $58s$ ($ i_{\text{pr}}=3$) spectrum, and barely observable for $59s$ ($ i_{\text{pr}}=5$). Because the integrated signal intensity around the peak centered
 at detuning $\epsilon_{\beta}/h$ reflects the overlap of the lattice eigenstate $\ket{\beta}$ with $\ket{i_{\text{pr}}}$ (Eq.\ \ref{eq:FGR}), this reveals that the edge states are localized on the weakly coupled boundary sites, with little contribution from  undressed bulk sites $58s$ ($i=3$) and $59s$ ($i=5$).  
 The widely split bulk states, however, have strong and approximately equal contributions  in spectra for detuning near the $58s$ and $59s$ undressed levels, and little contribution for detuning near $57s$.

The band structure as a function of strong tunneling rate $t_s$ [Fig. \ref{fig:Gallery}(c)] agrees with results from a direct diagonalization of Eq. \ref{eq:SSH} with $\delta_i=0$. For the configuration with strong tunneling to the boundary sites, the edge states vanish from the band structure [Fig. \ref{fig:Gallery}(d)].

\begin{figure}
  \includegraphics[scale = 0.7]{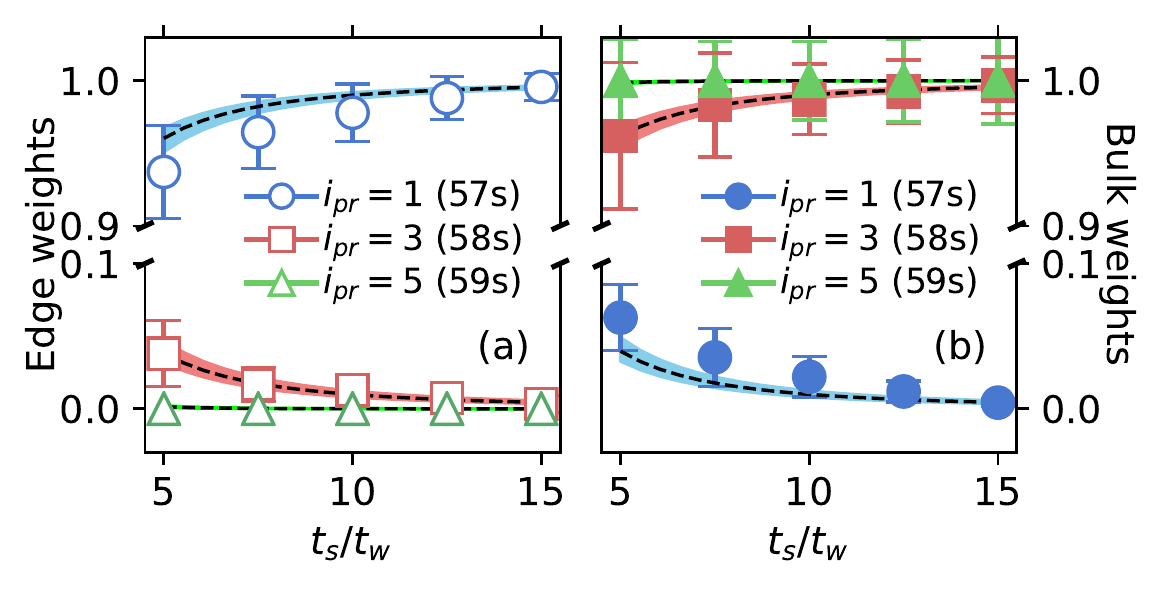}
 \caption{\label{fig:Trends}Synthetic-lattice-eigenstate decomposition  obtained from spectral-line areas [Fig.\ \ref{fig:Gallery}(a,b)]. (a) Fraction of the entire signal under the spectral features corresponding to the edge states  for  probe tuned near site $ i_{\text{pr}}$ (Rydberg level) indicated in the legend. The line is the sum of the squares of the calculated overlaps of the SSH edge eigenstates with the  $ i_{\text{pr}}$ site found from a direct diagonalization of Eq.~(\ref{eq:SSH}) with $\delta_i=0$.  (b) Fraction of the entire signal under the bulk state features and calculated  sum of squares of the overlaps of the SSH bulk eigenstates with the  $ i_{\text{pr}}$ site.
 }
\end{figure}

Diagonalization also provides the decomposition of each SSH eigenstate $\ket{\beta}$  upon the bare lattice sites, expressed in the factors $|\braket{\beta|i}|^2$. This can be compared with experimental measurements of  the fraction of the total 
spectral area in either the  edge or 
the bulk spectral features when probing the overlap with a specific lattice site ($ i_{\text{pr}}$) in spectra such as Fig. \ref{fig:Gallery}(a,b). Spectral area is determined by fitting  each of the three features in a spectrum with a sinc-squared lineshape corresponding to the $5\,\mu$s laser exposure time.
Figure \ref{fig:Trends} (left) shows that the experimentally measured edge-state fraction matches  $\sum_{\beta\in\text{edge}}|\braket{\beta| i_{\text{pr}}}|^2$, and Fig.\ \ref{fig:Trends} (right) does the same for the bulk contribution and $\sum_{\beta\in\text{bulk}}|\braket{\beta| i_{\text{pr}}}|^2$.
The width of the calculated line denotes 10\% variation in the Rabi frequencies.
For a given $t_s/t_w$, the edge-state measurements in Fig.\ \ref{fig:Trends} add to one, while the bulk-state measurements add to two. This reflects the fact that there are two edge states and four bulk states for this system, and half of the weight for the states in each group is in overlap with even lattice sites, which the photoexcitation probe does not detect.

\begin{figure}
  \includegraphics[scale = .72]{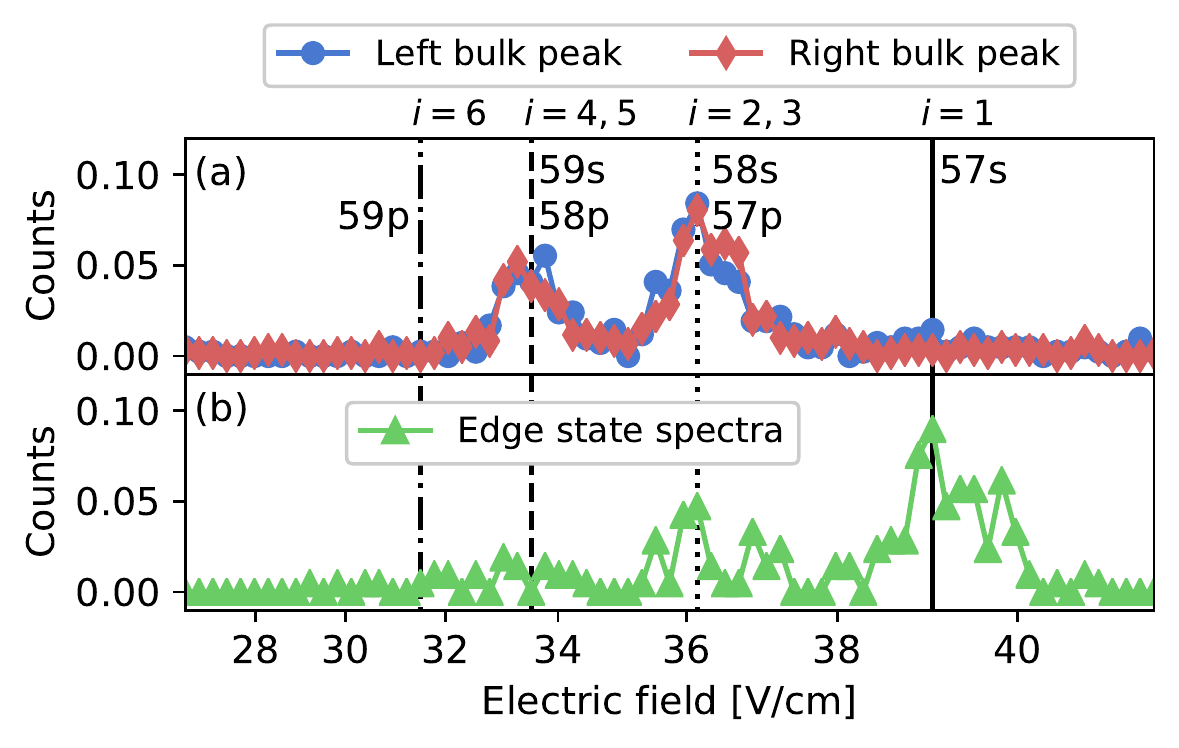}
 \caption{\label{fig:SFI} SFI signals for probe laser tuned near the $58s$ Rydberg level ($ i_{\text{pr}}=3$) for $t_s/t_w=5$ [Fig. \ref{fig:Gallery}(a)]. Vertical lines indicate ionization fields for bare Rydberg levels. Data points are evenly spaced in time. (a) For excitation to the left and right bulk-state peaks ($\Delta_{ i_{\text{pr}}}\approx \pm 500$\,kHz), the state excited is localized  on the bulk sites  of the synthetic lattice. (b) For excitation to the edge-state peak, ($\Delta_{ i_{\text{pr}}}\approx 0$) the state is localized on the $57s$ ($i=1$) boundary site. The small contribution to the signal at $i=2,3,4$, and $5$ is predominantly from the wings of the bulk-state peaks.}
\end{figure}

Because the spectral probe is only sensitive to  $ns$ contributions to the state vector (odd $i$), it cannot establish whether the edge states observed are localized on one boundary site or a superposition of both. To answer that question, we turn to SFI as a tool for site-population measurements in Rydberg-atom synthetic dimensions. 
For  Rydberg excitation near $58s$, corresponding to  $ i_{\text{pr}}=3$, and for $t_s/t_w=5$ [Fig.\ \ref{fig:Gallery}(a)], if  the detuning is set to resonance with the left or right bulk-state peaks ($\Delta_{ i_{\text{pr}}}\approx \pm 500$\,kHz), electrons are liberated at ionization fields for Rydberg levels corresponding entirely to bulk sites of the synthetic lattice ($i=2-5$) [Fig \ref{fig:SFI}(a)]. For laser detuning on the edge-state peak ($\Delta_{ i_{\text{pr}}}\approx 0$), signal arrives  at fields corresponding predominantly  to the $57s$ Rydberg state ($i=1$) [Fig \ref{fig:SFI}(b)]. This indicates  localization of the edge state on the boundary in general and more specifically on the single boundary site connected  to the ground-state by the two-photon excitation [Eq.~(\ref{eq:FGR})]. Integrals of SFI signals corresponding to each lattice site and for each spectral feature provide state decomposition that agrees with expectations as in Fig.\ \ref{fig:Trends}.

The pinning of the edge-state energy to $\Delta_{i_{pr}}=0$ is the defining feature of  TPS in the SSH model. It reflects an underlying chiral symmetry, which refers to the system's bipartite structure (even and odd sites), with all Hamiltonian matrix elements vanishing between sites of the same partition, including diagonal (on-site) matrix elements.  To investigate the robustness of the pinning of the edge-state energy, we probe the band structure in the presence of perturbations from the  SSH form.

\begin{figure}
  \includegraphics[scale = .6]{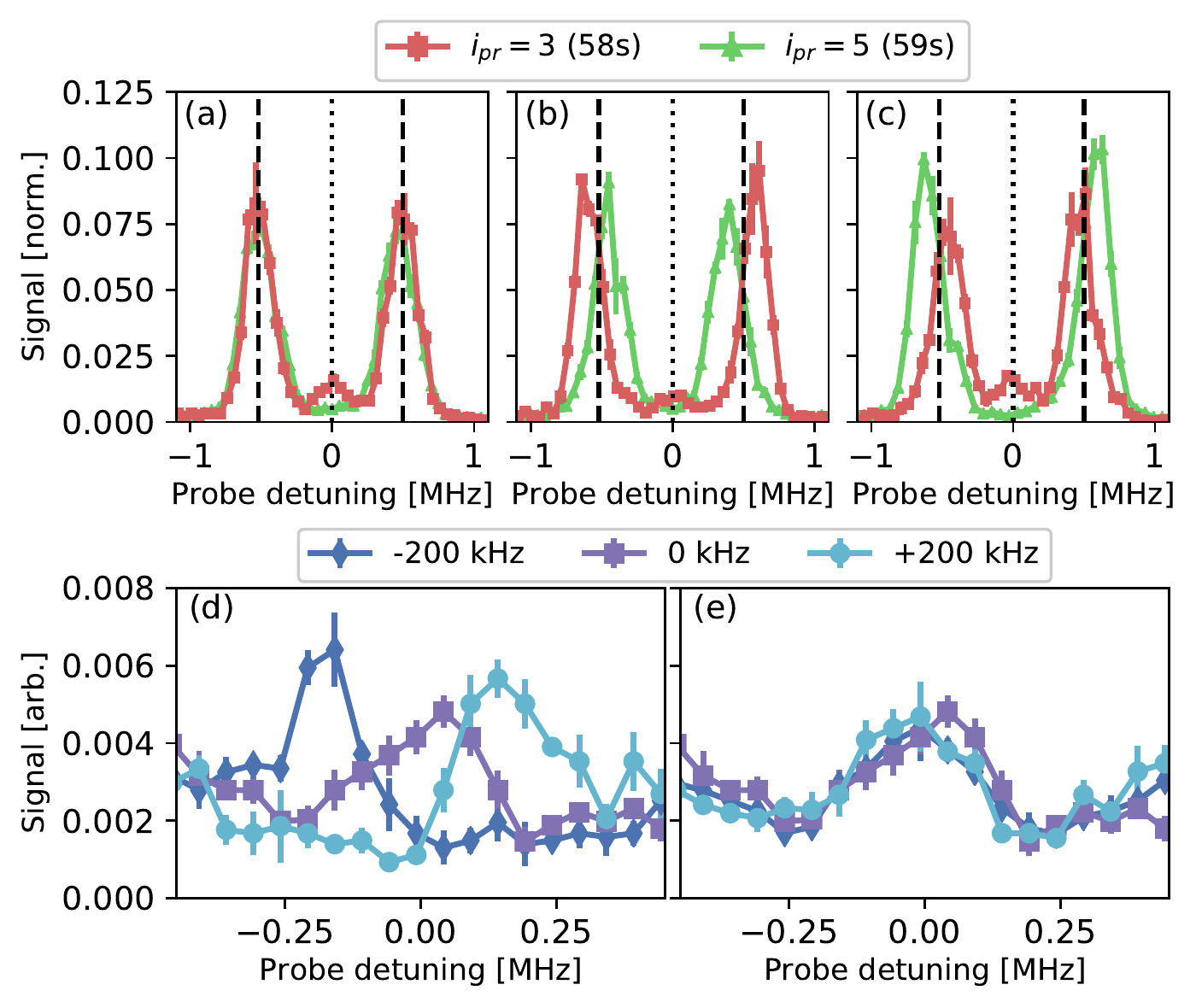}
 \caption{\label{fig:symmetry} Band structure with Hamiltonian perturbations. (a) SSH model: $\delta_i=0$ with balanced tunneling rates ($t_{2-3}$, $t_{4-5}$=$t_s^0$ and $t^0_s/t_w=5$). Lines mark positions of the bulk and edge peaks. (b) Strong tunneling rates are imbalanced to  $t_{2-3}=1.15 t^0_s $ and $t_{4-5}=0.85 t^0_s$ 
  for $t^0_s/t_w=5$. This perturbation respects chiral symmetry.  (c) Same as (b) but
 $t_{2-3}=0.85 t^0_s $ and $t_{4-5}=1.15 t^0_s$. 
 (d) Perturbation breaking chiral symmetry: tunneling rates are balanced as in the standard topological SSH configuration, with $t_s/t_w=10$, but the  frequency of the $i=1$ to $i=2$ ($57s$-$57p$) coupling is varied
 by the value given in the legend. 
 (e) Same as (d), but  the $i=5$ to $i=6$ ($59s$-$59p$) coupling frequency is varied.
}
\end{figure}

Figure \ref{fig:symmetry}(a,b,c) shows spectra for 
 $ i_{\text{pr}}=3$ ($58s$) and  $ i_{\text{pr}}=5$ ($59s$)  for  balanced ($t_{2-3}=t_{4-5}= t^0_s$) and imbalanced   
[$t_{2-3}=(1\pm0.15)t^0_s $ and $t_{4-5}=(1\mp0.15) t^0_s$] strong coupling.
Here, $t^0_s/t_w=5$ and $t_w=100\,$kHz. 
The bulk states are strongly affected by imbalance. 
With increased $t_{2-3}$ 
 [Fig.\ \ref{fig:symmetry}(b)], two bulk states that are more localized on the $i=3$  site  show increased splitting. With increased $t_{4-5}$ [Fig.\ \ref{fig:symmetry}(c)],  two bulk states that are more localized on the   $i=5$  site show increased splitting. 
The energy of the edge-state signal, however [in Fig.\ \ref{fig:symmetry}(b,c)], is
immune to this perturbation, which preserves the protecting chiral symmetry because the tunneling matrix elements only connect even and odd sites. 

Figure \ref{fig:symmetry}(d,e) shows how the energies of the edge states are affected by chiral-symmetry-breaking perturbations, in particular shifts of  on-site potentials (i.e. millimeter-wave coupling frequencies), such that $\delta_i\neq 0$ for some $i$. Spectra are recorded with the probe laser tuned near the  $58s$ level ($ i_{\text{pr}}=3$) for $t_s/t_w=10$.
For Fig. \ref{fig:symmetry}(d), the frequency of the $i=1$ to $i=2$ ($57s$-$57p$) coupling is varied, which shifts $\delta_1$, the on-site potential of the $i=1$ ($57s$) site in the synthetic lattice. This diagonal term in the Hamiltonian breaks the chiral symmetry, and the edge state signal shifts by an amount equal to the detuning from resonance. 
For Fig. \ref{fig:symmetry}(e), the frequency of the $i=5$ to $i=6$ ($59s$-$59p$) coupling is varied, shifting $\delta_6$, and the position of the edge-state signal remains unchanged. 
These results confirm that the edge state coupled to by the probe laser is  localized  on the $i=1$ ($57s$) boundary site. The orthogonal edge state is localized on $i=6$, with vanishing weight on odd sites, and this particular form of perturbation only affects the energy of one of the edge states.
In general we expect that any perturbation producing a Hamiltonian term that connects only even  sites to even  sites, or odd to odd, will break the chiral symmetry and shift  edge-state energies.

 We have demonstrated Rydberg-atom synthetic dimensions as a promising new platform for the study of quantum matter. The spectrum of photo-excitation to the synthetic lattice space formed by the manifold of coupled Rydberg levels provides the band structure and decomposition of the lattice eigenstates. SFI of the excited states provides an additional diagnostic of lattice-site populations with two-site resolution. TPS were observed in a six-site SSH model, and the measured band structure and eigenstate decomposition agree well with theory. Varying the detuning of the millimeter-wave fields that create tunneling between sites introduces on-site potentials, and this has been used to break the chiral symmetry of the SSH model and shift the energies of edge states away from the center of the bandgap at $\Delta_{i_{pr}}=0$.

The Rydberg-atom synthetic-dimension platform opens up exciting future directions. Additional millimeter-wave-coupling schemes and tunneling configurations are possible, such as two-photon transitions and transitions with larger changes in principal quantum number. This will enable creation of higher-dimensional synthetic lattices \cite{tsomokos2010, boada2012, jukic2013} and investigation of systems with non-trivial spatial \cite{Boada_2015} and band-structure \cite{bansil2016,qi2011} topologies and higher-order topological states \cite{bbh17}, for example.
Expanding the size of the synthetic space through the application of more millimeter-wave frequency components should be straightforward. Through control of millimeter-wave phases, tunneling phases around plaquettes and artificial gauge fields can be introduced \cite{gjo14}. This platform is also ideally suited for study of time dependent phenomena, such as Floquet-symmetry-protected states \cite{rha17}, non-equilibrium states \cite{mco19}, and wave-packet dynamics in synthetic space. 
The limit imposed by Rydberg-level decoherence needs further study, but  based on previous work demonstrating coherent manipulation of  Rydberg-level populations (e.g. \cite{blr15,sdf17}), decoherence should not  prevent observation of  the phenomena of interest. Tailored time variation of the electric-field ramp \cite{gbg18} may improve  site resolution of the SFI diagnostic. With systems of single Rydberg atoms in closely spaced optical tweezers \cite{ewl21,ssw21}  
with  long-range interactions in real space \cite{bbl16}, it should be possible to study interacting, many-body systems in synthetic lattices \cite{sgh18,stw19}. 

Acknowledgments: This research was supported by the AFOSR (FA9550-17-1-0366), NSF (PHY-1904294 and PHY-1848304), and the Robert A. Welch Foundation (C-0734, C-1844, and C-1872)

\bibliography{references_clean}

\end{document}